**Real-time Emergency Vehicle Event Detection Using Audio Data**

**Zubayer Islam**
Postdoctoral Scholar
Department of Civil Environmental and Construction Engineering
College of Engineering and Computer Science
University of Central Florida, Orlando, Florida 32826
Email: zubayer_islam@knights.ucf.edu

**Mohamed Abdel-Aty**
Pegasus Professor and Chair
Department of Civil Environmental and Construction Engineering
College of Engineering and Computer Science
University of Central Florida, Orlando, Florida 32826
Email: M.Aty@ucf.edu

**ABSTRACT**

In this work, we focus on detecting emergency vehicles using only audio data. Improved and quick detection can help in faster preemption of these vehicles at signalized intersections thereby reducing overall response time in case of emergencies. Important audio features were extracted from raw data and passed into extreme learning machines (ELM) for training. ELMs have been used in this work because of its simplicity and shorter run-time which can therefore be used for online learning. Recently, there have been many studies that focus on sound classification but most of the methods used are complex to train and implement. The results from this paper show that ELM can achieve similar performance with exceptionally shorter training times. The accuracy reported for ELM is about 97% for emergency vehicle detection (EVD).

**INTRODUCTION**

The response time of vehicles like ambulance, fire truck, and police vehicles are crucial in emergencies. In many cases, it depends on how quickly it can navigate through traffic to reach the destination. Historically, these vehicles have used lights and sirens to alert other road users. Therefore, detection of emergency vehicles using video and audio in real-time can aid in faster navigation especially through intersections. Some intersections are equipped with systems that can aid in emergency vehicle preemption with the help of a transmitter in the vehicles that can broadcast information to intersections. A receiver at the intersection is also required for this system. The use of audio data can aid in a much simpler system that could be cheaper and easier to implement on a larger scale. A mic can be used at the intersection which will be used to distinguish siren from emergency vehicles to provide the right of way. A multi-mic setup can also be used to understand the direction of the oncoming emergency vehicle.

Audio data has been studied in various fields for classification of sounds. In medical fields, there have been studies that focus on classifying heartbeat using audio data (Ankışhan, 2019; Raza et al., 2019). Speech recognition is being used widely in different smart devices (Anguera et al., 2014; Noda et al., 2015). The number of studies relating audio data classification for use in the field of transportation engineering is limited. To this end, this work proposes the detection of emergency vehicles using audio data.

Most of the methodologies that classify audio data require complex deep neural networks (Aytar et al., 2016; Korbar et al., 2018; Piczak, 2015a; Zhang et al., 2017) that can require extensive time to train. Especially in recent times when there is a lot of sensor data such methodologies can suffer from severe drawbacks. In addition, the different hyperparameters in the models can also delay the overall training time. We propose the use of Extreme Learning Machines (ELM) that has only one hidden layer and can be trained with only one epoch. The accuracy is also comparable to other complex neural networks in the literature while simultaneously reducing training time. These types of models can also aid in real-time online training mechanisms as well when no human supervision is required. As soon as new data is available, ELMs can be trained thereby keeping the model updated.

To train a model on acoustic data, it is important to extract the features from the audio recording. It is a two-step methodology, where at first features are extracted from the audio recording and then fed into a model for classification. In this study, the mel frequency cepstral coefficients (MFCCs) were extracted as well as the zero crossings. MFCCs describe the overall shape of the spectral envelope whereas zero-crossing rate (ZCR) is the rate at which the signal goes from positive to zero or negative to positive. To the best of the knowledge of the authors, ELM has not been used for any traffic application and can be highlighted as a contribution of the paper. Other contributions include the addition of ZCR as a feature, the use of data balancing techniques, and the comparison of various methods based on run-time for real-time online learning applications.

**RELATED WORK**

Siren sound detection has been studied to some extent with the help of collected data. Some studies focused on the algorithm while others suggested specific microcontroller-based prototypes for detection. Longest common subsequence was used by Liaw et al. (2013) while part-based models were used by Schröder et al. (2013) to analyze the spectro-temporal features of the sound data. A two-stage methodology was also studied where the first step is to detect abnormal sound and the second step is to perform classification (Marchegiani and Posner, 2017).

Support vector machines was analyzed by Carmel et al. (2017) on a small dataset. The reusability of such models for big data analytics remains questionable. Speech recognition-based systems were also used where at first features like MFCC were extracted and then trained on neural networks (Beritelli et al., 2006; Karpis, 2012). Convolutional neural networks were also studied by Tran and Tsai (2020) and impressive accuracy was reported with hybrid methodology. However, the model was trained on an unbalanced dataset which can often lead to inflated accuracy metrics. Sinusoidal models were also analyzed (Ellis, 2001) and features were also extracted from alarm sounds to capture alarms in noisy environments. Another detection method based on Doppler effect and fast fourier transform was proposed and implemented in a microcontroller (Miyazakia et al., 2013) but was computationally expensive, needing at least 8 seconds to make a classification. Other studies include pitch detection algorithm by Meucci et al. (2008) and linear prediction model by Park and Trevino (2013). Some low computational methods were also suggested based on analog circuits (Dobre et al., 2017; Dobre et al., 2015). In the current methodology, we have tried to reduce the complexity of training while improving upon the accuracy metrics. In the next section, a brief description of an ELM along with the proposed architecture is discussed.

**METHODS**

Extreme learning machine (Huang et al., 2006) is a feedforward single hidden layer neural network. It can be used both for classification and regression. The notable difference between this model and the other neural network models is that there is no iteration necessary to train ELM. The model can also adaptively select the number of nodes in the hidden layer. The output of the hidden layer is obtained by randomly selecting weights for the input layer. Generalized inverse of the hidden layer output matrix is used to calculate the bias and weights of the hidden layer.

A single layer feed-forward network (SLFN) with $L$ nodes can be mathematically modeled as

$$f_L(x) = \sum_{i=1}^{L} G_i(x, a_i, b_i) \cdot \beta_i \qquad a_i \in R^d, \beta_i \in R$$

Here, $a_i$ is the input weight vector, $b_i$ is the bias of the hidden layer and $\beta_i$ is the output layer weight. The function $G_i$ is the node activation function and typically takes the form:

$$G_i(x, a_i, b_i) = g(a_i \cdot x + b_i)$$

It has been shown that SLFNs can approximate a continuous target function (Huang et al., 2011) given that $x_i$ is the training data, $t_i$ is the labels and $L$ is the total hidden nodes in $\{(x_i, t_i) | x_i \in R^d, t_i \in R^m, i = 1, \ldots, N\}$. It has also been proven that randomly generated networks that have outputs reduced by least squares can maintain universal approximation capability. ELM has the objective function defined to reduce the norm of the output weights:

$$Minimize: ||\beta||_u^{\sigma_1} + \lambda ||H\beta - T||_v^{\sigma_2}$$

Here $H$ is the hidden layer output matrix and $T$ is the target matrix.

To summarize, firstly ELM learning algorithms randomly assign hidden node weights and biases $a_i$ and $b_i$. Secondly, the hidden layer output matrix is calculated. Finally, the output weight vectors were calculated with:

$$\beta = H^\dagger T$$

Here $H^\dagger$ is the Moore-Penrose generalized inverser of $H$ and $T = [t_1, \ldots, t_N]^T$.

The model proposed in the paper is summarized in **Figure 1**. At first, the audio data is segmented into fixed lengths since the number of input feature of the model need to be consistent. In the next step, data is balanced using SMOTE (Chawla et al., 2002). Two features were extracted such as MFCC and ZCR which were merged and then fed into ELM.

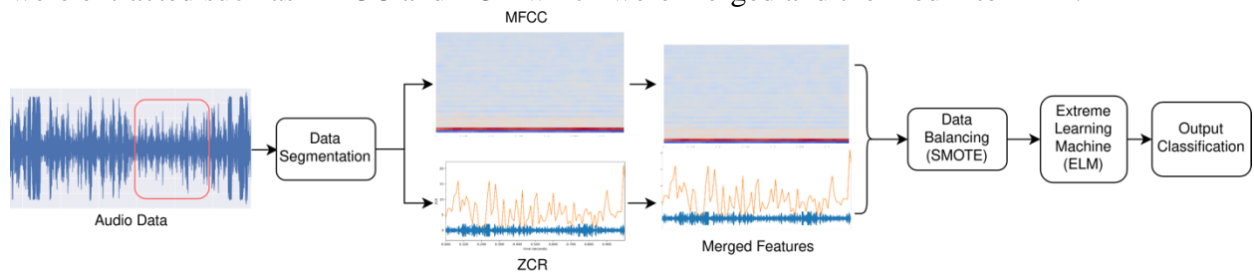

**Figure 1 Proposed Methodology**

### DATASET

We have used an open-source dataset to verify the effectiveness of the proposed methodology: ESC-50 (Piczak, 2015b). ESC-50 dataset contains 2000 audio clips. There are 50 labels in this dataset, and each has 40 samples. Since the application of this paper is focused on detecting siren sounds from the typical urban noises, we have removed some labels from ESC-50 such as animal sounds, natural sounds, etc. The classes that were used are summarized in **Table 1**.

**TABLE 1 Dataset Attributes**

| Dataset | ESC-50 |
| --- | --- |
| Class 1: Emergency Vehicle Sound | Siren |
| Class 2: Urban Sounds | Car horn, Engine, Train, Helicopter, Chainsaw, Airplane, Fireworks, Handsaw, Crying baby, Sneezing, Clapping, Coughing, Footsteps, Laughing, Rain, Wind |
| Average Clip Duration | 1 to 5 seconds |
| Total Duration | 2.8 hours |
| Sampling Rate | 44.1 kHz |

### Data Processing

To produce superior recognition performance, it is essential to capture important features. For each of the clips, we extracted two common features such as MFCC and ZCR. MFCC consists of two filters that can recognize frequencies both above and below 1 kHz. We have used the mean and standard deviation of 13 MFCCs from the input data. In addition, ZCR was also calculated which indicated the number of times a signal crosses the zero-amplitude line from positive to negative values or vice versa. The MFCC and ZCR (Das and Parekh, 2012) are calculated by the equations given below, where $f$ is the frequency and $sgn(x_i)$ indicated the sign of the $i$-th sample of the audio data.

$$MFCC = 2595 \log\left(1 + \frac{f}{700}\right),$$

$$ZCR = \sum_{i=1}^{\omega} \frac{|sgn(x_i) - sgn(x_{i-1})|}{2}$$

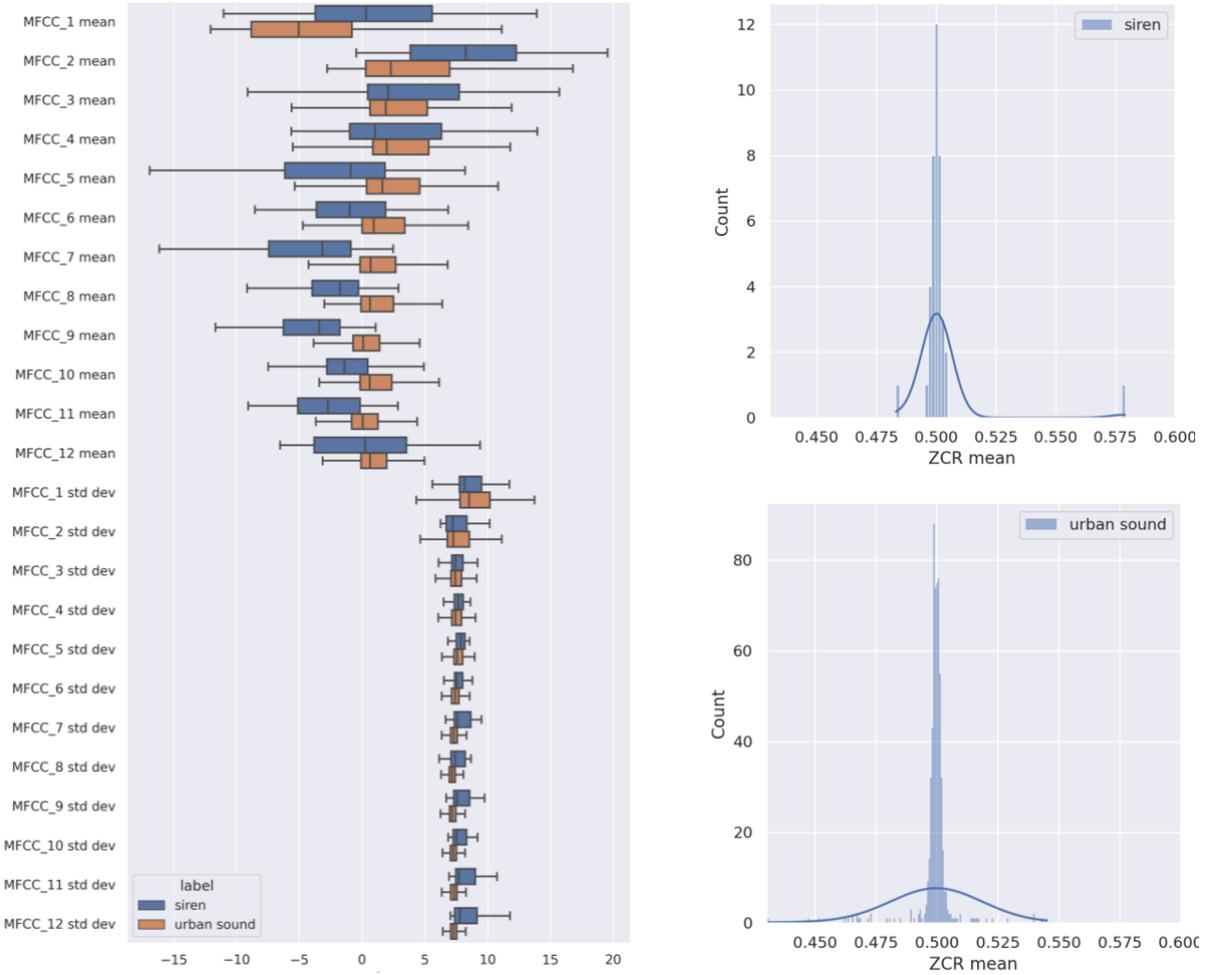

**Figure 2 MFCC and ZCR comparison between the two classes**

The comparative analysis of the features between the two classes of the dataset is presented in **Figure 2**. Here we can see that of the 13 MFCC values, 5 to 10 have different distribution for siren and urban sounds and can be highlighted as the notable features to differentiate the classes. The distribution of the ZCR also shows that siren noise can be distinguished from urban noise. After feature extraction, four of the folds of the data were used to train the model while testing was performed on the remaining fold.

**RESULTS**

The results are discussed with two metrics: accuracy and run-time. Five different models were trained alongside ELM such as K-Nearest Neighbors (KNN), Random Forest (RF), Support Vector Machine (SVM), Multilayer Perceptron (MLP) and Convolutional Neural Network (CNN). The time required by each model to train on the dataset is shown in **Table 2**. The run-time of ELM is better than the next-best performing model KNN by 10 times. It is also evident that complex neural net models such as MLP and CNN have the worst run-time (about 4000 times slower than ELM). It is also important to mention that the MLP and CNN models used for this study had a simple structure. MLP had three hidden layers with 25, 16, 12 neurons while CNN had two 1D convolutional layers of 20 and 16 filters followed by a 16-neuron hidden layer. Therefore, any complex system than this would also perform worse in terms of run-time.

Next, we evaluate the accuracy of the models as shown in **Table 3**. The best performing model here is ELM with an accuracy of 94.5%. SVM has the nearest accuracy metrics, but it is at the expense of a run-time 48 times slower than ELM. As we have used rudimentary level MLP and CNN, the accuracy is poor. As shown in several studies (Tran and Tsai, 2020) that proper tuning of these models would aid better accuracy but the run-time would be worse than those shown in **Table 2**.

**TABLE 2 Run Time Comparison (in milliseconds)**

|        | KNN  | RF     | SVM   | MLP     | CNN    | ELM  |
|--------|------|--------|-------|---------|--------|------|
| Fold 1 | 3.36 | 977.8  | 12.42 | 1043.8  | 1084.0 | 0.26 |
| Fold 2 | 2.49 | 1000.6 | 11.73 | 1044.3  | 1105.1 | 0.24 |
| Fold 3 | 2.46 | 951.6  | 9.44  | 1061.5  | 1101.9 | 0.26 |
| Fold 4 | 2.44 | 970.2  | 11.67 | 1082.9  | 1110.3 | 0.24 |
| Fold 5 | 2.43 | 1024.9 | 15.01 | 1049.1  | 1105.7 | 0.24 |
| Overall| 2.64 | 985.02 | 12.05 | 1056.32 | 1101.4 | **0.25** |

**TABLE 3 Accuracy Comparison (%)**

|        | KNN  | RF   | SVM  | MLP  | CNN  | ELM   |
|--------|------|------|------|------|------|-------|
| Fold 1 | 92.1 | 72.6 | 96.0 | 32.8 | 28.5 | 97.05 |
| Fold 2 | 95.3 | 91.0 | 93.3 | 32.8 | 44.5 | 94.85 |
| Fold 3 | 83.9 | 62.8 | 85.1 | 40.2 | 40.6 | 95.58 |
| Fold 4 | 58.5 | 61.7 | 73.4 | 33.9 | 40.2 | 91.91 |
| Fold 5 | 96.4 | 84.3 | 98.0 | 47.2 | 35.1 | 93.38 |
| Overall| 85.3 | 74.5 | 89.2 | 37.4 | 37.8 | **94.55** |

Theoretically, ELMs should be able to classify with perfect accuracy if the number of hidden neurons is increased. In **Table 4**, the accuracy and run-time are presented as the number of hidden neurons is varied from 10 to 10000. The best accuracy is reported at 10000 neurons, but the run-time is worse than the baseline of 10 neurons. Even then, the run-time is better than the MLP and CNN models. As the number of neurons is increased the accuracy does not change dramatically but with slight improvements. Therefore, the conclusion is that for quick and reliable emergency vehicle siren sound detection, it is reasonable to use 10 or 100 neurons in an ELM.

Lastly, the proposed methodology is compared with the existing studies. **Table 5** shows that our proposed model performs better than most studies (Liaw et al., 2013; Marchegiani and Posner, 2017; Schröder et al., 2013). We would also like to note that it performs almost equally with the recently proposed method SirenNet (Tran and Tsai, 2020). As we explained in Table 2 that the run-time of rudimentary CNN methods is significantly worse than ELMs, it would be worse for complicated models such as SirenNet with only marginal improvement in accuracy.

**TABLE 4 Accuracy and run-time variations with neurons in hidden Layer**

| # of neurons in hidden layer | 10 | | 100 | | 1000 | | 10000 | |
|---|---|---|---|---|---|---|---|---|
| | Accuracy | Time | Accuracy | Time | Accuracy | Time | Accuracy | Time |
| Fold 1 | 97.0 | 0.26 | 97.0 | 18.9 | 93.3 | 203.2 | 99.2 | 829.1 |
| Fold 2 | 94.8 | 0.24 | 95.5 | 18.9 | 91.1 | 138.9 | 97.0 | 766.8 |
| Fold 3 | 95.5 | 0.26 | 96.3 | 16.1 | 94.8 | 125.7 | 96.3 | 872.9 |
| Fold 4 | 91.9 | 0.24 | 94.1 | 16.0 | 88.2 | 130.0 | 94.1 | 778.4 |
| Fold 5 | 93.3 | 0.24 | 95.5 | 6.5 | 93.3 | 134.8 | 97.0 | 893.0 |
| Overall | 94.5 | **0.25** | 95.68 | 15.28 | 92.14 | 146.52 | **96.72** | 828.04 |

**TABLE 5 Comparison with studies in the literature**

| Study | Features | Method | Accuracy |
|---|---|---|---|
| Schröder et al. (2013) | MFCC, Spectrogram | Part based model, HMM | 74% to 86% |
| Liaw et al. (2013) | Longest Common Subsequence | Comparison | 85% |
| Marchegiani and Posner (2017) | Empirical Binary Mask, Gammatonegrams | KNN | 62% to 83% |
| Tran and Tsai (2020) | MFCC, Spectrogram | CNN | 92% to 98% |
| Proposed Methodology | MFCC, ZCR | ELM | **92% to 97%** |

**CONCLUSIONS**

In this study, we proposed the use of Extreme Learning Machines in the field of audio data classification. Features such as MFCC and ZCR were extracted from the raw data. The data was also balanced with SMOTE to avoid inflated accuracies. The main motivation behind using a simplistic model such as ELM was to illustrate the huge improvements in run-time that is possible without compromising accuracy. With ELMs we have achieved accuracy up to 97% that is comparable to the current state of the art while dramatically reducing the training time. ELMs are 10 times faster than KNN and over 4000 times faster than state of the art CNN or MLP models. These types of models will be especially suitable for online machine learning where a model is constantly updated with newer data rather than relying on the previous data that might be rendered obsolete. We also have used ZCR as a feature that has not been studied in this area to the best of our knowledge.

Sound sensing can bring a new dimension in traffic engineering where we mostly rely on camera, radar, or loop detectors. These types of data can aid in finding special events that cannot be otherwise understood. For example, a tire skidding sound can only be captured with the help of audio data and that can be an indicator of harsh driving event. Camera, radar, or loop detectors cannot report such special safety events. Certain weather conditions such as rain and thunderstorm can also be picked up by audio sources. Future work can extend the proposed methodology for urban sound classification in general. Moreover, anomaly or special event detection in roadways can also be studied.

**ACKNOWLEDGMENTS**
The authors acknowledge the funding of the Florida Department of Transportation (FDOT) for this study.

**AUTHOR CONTRIBUTIONS**
The authors confirm contribution to the paper as follows: study conception and design: Zubayer Islam, Mohamed Abdel-Aty; analysis and interpretation of results: Zubayer Islam, Mohamed Abdel-Aty; draft manuscript preparation: Zubayer Islam. All authors reviewed the results and approved the final version of the manuscript.

**CONFLICT OF INTEREST**
On behalf of all authors, the corresponding author states that there is no conflict of interest.